\documentstyle[psfig,preprint,pre,aps,eqsecnum]{revtex}
\tightenlines
\begin{document}
\bibliographystyle{normal}

\draft
\title{Periodic forcing in viscous fingering of a nematic liquid crystal}

\author{R. Folch$^{1,2,*}$, T. T\'oth-Katona$^{3,\dag}$, \'A. Buka$^3$, 
J. Casademunt$^1$ and A. Hern\'andez-Machado$^{1,2}$}

\address{
$^1$ Departament d'Estructura i Constituents de la Mat\`eria, 
Universitat de Barcelona, \\
Avinguda Diagonal, 647, E-08028-Barcelona, Spain \\
$^2$Groupe de Physique des Solides, UMR 7588, 
CNRS / Universit\'es Paris VII et Paris VI, \\
Tour 23, 2 place Jussieu, F-75251 Paris Cedex 05, France\\
$^3$ Research Institute for Solid State Physics and Optics, 
Hungarian Academy of Sciences, \\ 
H-1525 Budapest, P.O.B.49, Hungary \\
$^*$ e-mail addresses: rf@pmc.polytechnique.fr, roger@gps.jussieu.fr, 
roger@ecm.ub.es.\\
$^\dag$ Present address: Physics Department, Kent State University,
P.O.Box 5190, Kent OH 44242
}

\date{\today}
\maketitle
\begin{abstract}
We study viscous fingering of an air -- nematic interface 
in a radial Hele-Shaw cell when periodically switching on and off an electric 
field, which reorients the nematic and thus changes its viscosity, as well as
the surface tension and its anisotropy (mainly enforced by a single groove
in the cell). We observe undulations at the sides of the fingers
which correlate with the switching frequency 
and with tip oscillations which give maximal velocity to smallest
curvatures. These lateral undulations appear to be decoupled from
spontaneous (noise-induced) side branching. 
We conclude that the lateral undulations are generated by
successive relaxations between two limiting finger widths. 
The change between these two selected pattern scales is
mainly due to the change in the anisotropy.
This scenario is confirmed by numerical simulations 
in the channel geometry, using a
phase-field model for anisotropic viscous fingering. 
\end{abstract}

\pacs{PACS number(s): 47.54.+r, 47.20.Ma, 61.30.-v, 47.20.Hw}

%\begin{multicols}{2}
\section{Introduction}
\label{intro}

Interfacial instabilities constitute a diverse domain in non-equilibrium
pattern formation, with examples ranging from biology (e.g. bacterial growth)
to mathematics (Stefan problems), passing by physical systems 
as flows in porous media, solidification, electrodeposition or flame 
propagation \cite{pelce}. 

Progress in this domain has usually been made by 
studying prototype systems as solidification 
or viscous fingering. The latter deals with the destabilization of the 
interface between two immiscible fluids when the more 
viscous fluid is displaced by the less viscous one, 
which is either injected at an end of a channel-shaped cell (channel geometry) 
or from the center of the cell (radial geometry) ---for a review see 
\cite{mccloud1}. This initial destabilization leads to the formation of fingers
in both geometries, which finally restabilize into a single stationary finger 
in the channel one. However, a sufficient amount of noise may cause this single
finger to tip-split. In contrast, in the isotropic, radial cell, fingers do not
stabilize, but repeatedly tip-split to form more and more fingers 
\cite{pater1}.

An external perturbation, however, can dramatically change the fingers
and can even suppress the tip splitting in both geometries. 
A bubble of gas trapped just 
in front of an advancing finger causes tip stabilization and (eventually)
intensive and very regular side branching both 
in the radial \cite{couder1} and channel \cite{couder3} geometries. 
Engraving a grid on one of the plates of the radial cell introduces an 
anisotropy, which, if strong enough, also inhibits tip splitting and produces
dendrites and faceted structures, resulting in a rich morphology diagram 
\cite{benj9,benj4}. Different etched lattices 
give a variety of highly branched structures whose symmetry depends on that 
of the lattice when the anisotropy it introduces is strong enough 
\cite{chen}. 
The replacement of the grid by a set of parallel grooves 
\cite{horvath1} has produced an even a more complicated morphology diagram 
than that presented in \cite{benj9}.
With a single groove running from the center 
to the edge of the cell the tips split in all  
directions except that of the groove \cite{matsu1}, in which a much faster 
growing dendritic structure is observed, and the 
whole pattern is very similar to that reported in \cite{couder1}. 
An intrinsic anisotropy, such as that of a liquid crystal used as the
more viscous fluid, has also been shown to stabilize
the tips 
and yield growth with side branches \cite{buka96,lc}.
All these experiments have demonstrated that different kinds of anisotropy
affect viscous fingering as that of the surface tension does for dendritic
crystal growth, i.e., stabilizing finger tips, so that, 
if the natural noise is strong enough, destabilization of the finger takes 
place only at its sides in the form of side branches.

In the channel geometry, Rabaud {\it et al.} have taken advantage of the fact
that fingers remain stable up to higher capillary numbers once the 
introduced anisotropy
has suppressed tip splitting to artificially 
induce side branching by means
of an external perturbation \cite{rabaud}. 
This should enable one to study the side branching in a more controlled way,
and also the coupling between the perturbation and the branching dynamics.
They obtain side branches using a localized disturbance, namely
a knot on the thread which provides the anisotropy. 
Pressure modulation also causes side branching in the
case of a thread, since, according to them, it mainly induces localized
initial disturbances near the intersection of the interface with the 
thread. In contrast, in the case of two opposite grooves
in the middle of the channel, the lateral waves caused by such 
sinusoidal pressure oscillations are symmetrical, and, most significantly,
of limited amplitude.

This brings us to the fundamental problem of the general response of a 
pattern-forming interface 
to the {\em non-localized} periodic forcing of its dynamics. 
We study this response and the possible 
formation of lateral waves in an air finger invading
a liquid crystal in the radial geometry, where the boundary conditions 
would not limit their amplitude, when periodically forcing the system
by a modulated electric field perpendicular 
to the cell. A single groove 
running over the injection point stabilizes the finger 
tips in its direction. 
The nematic director tends to align with the electric field 
when this is switched on, and returns roughly
to the cell plane when this is switched off. 
The flow properties depend on the orientation of the 
director, so that we expect to change the control 
parameters of the dynamics whenever we switch the 
field on or off.
The use of a square wave for
the amplitude of the electric field (switching it on and off instantly)
enables us to observe the relaxation of the pattern to a
parameter quench.  
 
We find the tip radius to relax very quickly to two different
values when the field is switched on and off, and that this pulsating
tip induces symmetrical lateral undulations. 
Finally, we explain these lateral
undulations as the trace of a periodic change in the selected tip radius, 
caused mainly by the change in the effective anisotropy due to the interplay
between the liquid crystal and the groove. 
Back to the channel geometry, we confirm this scenario by numerically
integrating a phase-field model for viscous fingering \cite{pf} in
which the anisotropy is switched between two different values. 
Here, the alternate
relaxation towards two different selected pattern scales is particularly clear,
since the symmetrical lateral
undulations saturate, as the finger oscillates between two different selected
widths. 

This mechanism might be relevant to experiments in which similar observations
have been made. For instance, to the case of symmetrical undulations at the 
sides of a finger perturbed by a bubble on its tip, in which the tip curvature 
oscillates \cite{couder1,couder3,rabaud} and the lateral undulations in the 
channel geometry also lie between two well-defined asymptotic widths, with a 
Saffman--Taylor finger as outer envelop \cite{couder3,rabaud}. Another example
could be the 
sinusoidal modulation of the injection pressure in fingers grown with two 
parallel grooves, which also displayed symmetrical lateral waves of limited 
amplitude \cite{rabaud}.

The rest of the paper is organized as follows: 
In Sec. \ref{secexp} we present the experimental setup and observations. 
Sec. \ref{sectheory}
then introduces and exploits the 
theoretical framework within which we explain these experimental results,
and Sec. \ref{secnumerical},
the numerical method for checking the out-coming hypothesis in the
channel geometry. The conclusions reached are
summarized in Sec. \ref{secconclusions}.

\section{Experimental setup and results}
\label{secexp}

The experiments were performed in a radial Hele-Shaw cell. This was assembled
from two glass plates coated with a conducting layer of ${\rm SnO_2}$, which served
as electrode. The bottom plate, of dimensions $160mm\times 160mm$ and thickness
$5.5mm$, had a hole of $1mm$ diameter in the center as an inlet for the air. 
On the coated face of the upper plate ($140mm\times 140mm$ and thickness $3.1mm$) 
we engraved a groove following its diagonal.
The plates were separated by
$d = 0.32mm$ or $d = 0.19mm$ thick spacers. The inner faces of the plates
corresponded to the coated ones, so that the electrodes directly faced each 
other, with no glass in between.

We applied an AC electric field $E$ of frequency $1kHz$ perpendicular to the 
plates, and switched it on/off with a frequency $\nu$. 
The `semiperiods' during which $E$ was on ($t_{on}$) and off were unequal.
Their ratio was chosen so that fingers advanced a similar distance in each
`semiperiod', which resulted in more apparent effects. Thus,
a filling coefficient of $\xi = t_{on} \nu = 0.67 \pm 0.03$ 
was found to be convenient, and it was used in all the experiments presented
here. 

Initially, the cell was filled with the commercial liquid crystal mixture 
RO-TN-430 (La Roche), with positive dielectric anisotropy,
$\varepsilon_a = \varepsilon_{\parallel}-\varepsilon_{\perp} = 17.6$ 
($\varepsilon_{\parallel}$ and $\varepsilon_{\perp}$ 
are the components of the uniaxial dielectric tensor parallel and 
perpendicular to the director, respectively), 
and a broad temperature range of the nematic phase,
from $T_m = -10^{\circ}C$ up to $T_{N \rightarrow I} = 70^{\circ}C$ 
(experiments were performed at room temperature $T = 23^{\circ}C$). 
The mixture was doped with dichroic blue dye D16 (BDH) 
in order to enhance the contrast at the air--nematic interface.

Then, after being filtered, air was injected through the hole of the bottom 
plate at an excess pressure $p_e$, regulated by a ported 
precision regulator (Norgren 11-818-100) with an accuracy of 
$\pm 0.03bar$, and further 
decreased and stabilized by a unit for pressure reduction. 
The path of the air was regulated by two 3-path solenoid valves, and 
$p_e$ was measured with a precision pressure meter (Watson$\&$Smith), 
with an accuracy of $\pm 1mbar$.

As the air displaced the liquid crystal, a camera
recorded the growth process, and images were fed into a PC for 
digital analysis, with a spatial resolution of 512 pixels $\times$
512 pixels and 
a 256 gray scale for each pixel. With the magnification used, 
a spatial resolution of $(0.241mm \times 0.166mm)/\rm pixel$ 
was determined. 

Experimental results are presented in Figs. 1-4.
In all cases, two air fingers whose tips do not split grow along the groove,
at each side of the injection hole. Two much slower air bumps
form at each side of the groove and perpendicular to it (Figs. 1, 3 and 4), 
and their tips can split [Fig. 4(a)].

At low excess pressures ($p_e=5mbar$, Figs. 1-3) 
the two stable viscous fingers along the groove
do not show any lateral undulations with ($E$ on) or without ($E$ off) 
an AC field kept constant [Figs. 1(a) and 1(b), $d=320\mu m$, and 
3(a) and 3(b), $d=190\mu m$]. 
However, fingers grown with the field [Figs. 1(b) and 3(b)] are thinner and
slower (compare the times indicated in the captions) than their 
analogues grown without it [Figs. 1(a) and 3(a), respectively].

If one then periodically switches on and off the field (modulated $E$), 
the tips undergo successive curvature changes 
which induce formation of undulations at the sides of the two stable fingers
in a strong correlation with 
the switching frequency $\nu$, 
as shown for two different ones in Figs. 1(c) and 1(d),
where interfaces are displayed each time the field is switched on/off. 
Note that the maxima and minima of these lateral undulations
in those figures 
roughly overlap with the profiles of the 
fingers grown in the same conditions but
with the field kept off [Fig. 1(a)] and on [Fig. 1(b)] respectively. 

%%%%%%%%%%%%%%%%%%%%%%%%% FIG. 1 %%%%%%%%%%%%%%%%%%%%%%%%%%%%%%%%%%%%

Similarly to the solidification of a nematic liquid crystal into a smectic B
reported in \cite{borzs99}, 
above a certain switching frequency $\nu _c$ the main fingers 
show no lateral undulations [Fig. 1(e)]. Their widths are then intermediate, 
lying between those of Fig. 1(a) ($E$ off) and Fig. 1(b) ($E$ on). 

Fig. 2 plots the position of the tip of the fingers in Fig. 1
vs. time. Here it is apparent that fingers grow 
faster with $E$ off [wider fingers of Fig. 1(a), empty circles] than with 
$E$ on [thinner fingers of Fig. 1(b), filled circles], but also that the 
oscillations of the tip curvature in time when periodically switching on/off 
the field of Figs. 1(c) and 1(d) 
are accompanied by tip velocity oscillations (squares
and triangles, respectively). In each oscillation, when the field
is off (empty squares and triangles) and on (filled ones)
the velocity roughly attains the values
obtained for fingers grown with the field kept off (empty circles) and on 
(filled circles) all the time, respectively. Even for $\nu > \nu _c$, 
when no lateral undulations occur [Fig. 1(e)], the velocity 
increases and decreases significantly when 
the field is switched off
(empty diamonds) and on (filled diamonds), respectively,
although it is not clear whether
it attains the same values than for a constant field. 

%%%%%%%%%%%%%%%%%%%% FIG. 2 %%%%%%%%%%%%%%%%%%%%%%%%%%%%%%%%%%%%%%%%%

With the same $p_e=5mbar$ but a smaller cell gap $d=190\mu m$ (Fig. 3), 
all the previous qualitative observations are reproduced, 
but now all fingers are narrower [than in Fig. 1, compare Fig. 3(a) with 1(a)
and 3(b) with 1(b)]. 

%%%%%%%%%%%%%%%%%%%%% FIG. 3 %%%%%%%%%%%%%%%%%%%%%%%%%%%%%%%%%%%%%%%%%

At higher excess pressure ($p_e=22mbar$) and the same cell gap $d=190\mu m$ 
as in Fig. 3, the fingers along the grooves
show a weak, uncorrelated lateral noise both 
with $E$ off and on [Figs. 4(a) and 4(b)], whereas their overall widths 
roughly stay the same [compare with Figs. 3(a) and 3(b) respectively].
When periodically switching the field on and off, the lateral undulations
correlated with the switching frequency of Figs. 1(c) and 1(d) reappear, 
but now superimposed to and apparently
decoupled from the uncorrelated lateral noise [Fig. 4(c)].

%%%%%%%%%%%%%%%%%%%%%%%%% FIG. 4 %%%%%%%%%%%%%%%%%%%%%%%%%%%%%%%%%%%%%%%

Also at this higher excess pressure $p_e=22mbar$ do the regular lateral
undulations disappear above a certain frequency $\nu _c$ and leave the bare
uncorrelated noise of Figs. 4(a) and 4(b). This upper frequency 
$\nu _c$ shows a roughly linear dependence on the excess pressure $p_e$.

\section{Theoretical discussion}
\label{sectheory}

We now present a possible simplified theoretical framework to explain 
the experimental observations. 

The shear viscosity of a nematic liquid crystal
flowing in a planar cell depends on the orientation of its director:
The highest viscosity is achieved with the director perpendicular to the cell 
(homeotropic alignment). With the director lying on the cell plane
(planar alignment) the viscosity is lower and anisotropic:
higher with the director perpendicular than parallel to the flow.

The director and the velocity fields 
are coupled by nonlinear nematohydrodynamic equations ---see e.g. 
\cite{pleiner1,dubois1}. Thus, when the electric field is off, the flow
forces the director to be roughly in the plane of the cell 
(planar alignment case).
Moreover, the director tends to align itself with the
flow velocity in a certain pressure range. 
The latter, together with the
mentioned anisotropy of the viscosity in the cell plane respect to
the director orientation,
results in a viscosity which depends on the velocity direction. 
This causes the viscosity to be non-uniform and
anisotropic respect to the direction of the flow. This anisotropy turns
out to be the most important effect, since, 
if strong enough, it stabilizes the finger tips, 
thus
switching from a tip-splitting to a side-branching mode \cite{buka96}. 
This can
be understood by mapping this anisotropy in the viscosity to an effective
anisotropy in the surface tension \cite{lc}. 

However, experiments
performed without any groove found {\em no}
regime for which
this anisotropy was
strong enough to clearly stabilize the 
finger tips for the liquid crystal mixture used here \cite{toth00}, whereas
the introduction of the groove did stabilize them. 
Therefore, as a first approximation, we will neglect the anisotropic effect
of the director alignment in front of that of the groove: On the one hand,
we will consider this planar alignment case to have a unique, uniform and 
isotropic average viscosity; on the other hand,
we will not consider the effective
anisotropy in the surface tension coming from that in the viscosity,
but only the stronger anisotropy introduced
by the groove. Actually considering both of them does not change
qualitatively the simulation results. 

An AC electric field also exerts a torque on the director. For a
liquid crystal with positive dielectric anisotropy, 
$\varepsilon_a > 0$ as ours, and a field 
perpendicular to the plates,
the electric torque competes with the shear one, trying to align the director 
with the field, i.e., perpendicular to the plates (homeotropic alignment). 
Therefore, the viscosity with $E$ on should now be even more
isotropic in the shear plane. 

Consequently, both with $E$ off and on we will consider a constant, isotropic
viscosity. The only difference between the viscosities with and without electric
field will be that the viscosity with $E$ on should always be larger than with
$E$ off, since the field favors the homeotropic alignment. Note that this 
inequality of the viscosities with and without field will hold even for an
incomplete alignment.

Thus, the theoretical framework will be that of the standard
viscous fingering equations, except for the dimensionless surface tension 
$B_0\equiv \sigma/(p_e l_c)$ (with $\sigma$ the surface tension and $l_c$ an
arbitrary length scale), which will read
\begin{equation}
\label{eq:anisotropy}
B=B_0 \left(1-\alpha \cos^2\phi\right),
\end{equation}
where $\phi$ is the angle between the single groove used 
in the experiments
and the normal to the interface, and $\alpha$ represents the 
two-fold anisotropy induced by the
groove. 
Grooves and grids have
usually been modeled by such an anisotropy in the surface tension
(see, e.g., Ref. \cite{benjacob}). 
This represents a strong simplification,
but we do not expect it to affect the conclusions of this paper in any
fundamental way. 

Each time we switch $E$ on or off, we change the director orientation,
and thus some physical
parameters of the model, which should result in a change in its
dimensionless control parameters, namely $B_0$ and $\alpha$, or in the
time scale of the dynamics, $12\mu l_c^2/(d^2 p_e)$, where $\mu$ is
the viscosity. Since we always
switch $E$ on or off instantly, the adimensionalization leading to this set
of control parameters remains valid, even if their value is periodically
switched.
Before using evidence from the experiments, let us discuss for clarity
how can these control parameters and time scale be expected to change 
{\it a priori}.

The time scale $12\mu l_c^2/(d^2 p_e)$ can change only through a change in 
the viscosity $\mu$. Indeed, the change in the time
scale was measured directly for the same mixture from the growth of
a circular interface in the absence of grooves with $E$ on and off,
and it was found to be a factor $3.7$ 
slower with $E$ on \cite{toth00}. We thus know that the viscosity
is 3.7 times larger with $E$ on. 
(Note that this does not alter $B_0=\sigma/(p_e l_c)$,
since it does not depend on $\mu$ for an experiment at constant excess 
pressure, whereas it does for the constant injection rate case, for which its
definition is different). $B_0=\sigma/(p_e l_c)$ could only be altered by a
variation in the surface tension $\sigma$. Such a variation has been measured
for several liquid crystals, and $\sigma$ has been found to be 20--50\%
smaller with the director parallel to the air-nematic interface (roughly
our $E$ on case) than perpendicular to it \cite{tsvet1} (closer to $E$ off).

As for the anisotropy $\alpha$ due to the groove, it could be changed by the
following effect: The director inside the groove might
keep the planar alignment to some extent even with $E$ on, 
since the conducting layer was removed from the etched region
when engraving the groove. In that case, 
the viscosity would be lower inside than outside the groove with $E$ on, thus 
reinforcing the mobility enhancement of the groove itself (higher gap $d$), and
therefore increasing the effect of the groove (the anisotropy $\alpha$ in our
model). 

Now, consider the experimental results reported in the previous section,
and first of all, those for a field kept either on or off during all the
experiment.
The main fingers were found to be slower and thinner with $E$ on.
Their smaller velocity is
explained by the increase in the time scale of the dynamics due to that in
the viscosity, whereas their smaller width should be understood as a decrease 
in the selected finger tip radius 
for a given length of the finger,
which results in a visually overall thinner finger. (Thus we will talk about
thinner and wider fingers to refer to larger and smaller tip curvatures at
a given finger length, respectively). 
In our model, this decrease in the selected length scale
could be due to either a decrease in the 
dimensionless surface tension $B_0$ or an increase in its anisotropy $\alpha$.

To check the two possibilities, we varied $B_0$ and $\alpha$ by means other
than reorienting the director. In order to increase $\alpha$, we decreased
the cell gap $d$, which is the standard way of increasing the effect of a 
groove or grid \cite{benj9,benjacob}, and which does not affect anything else
in our model but the time scale. 
As mentioned in the previous section, the fingers do narrow. 
Note also that very similar widths are obtained either by switching on the
field [Fig. 1(b)] or by increasing $\alpha$ through a decrease in $d$
[Fig. 3(a)]. This proves that the observed finger narrowing 
when switching on the field can be caused by an increase in $\alpha$.
Consistently with this hypothesis, if $\alpha$ is further
increased by switching on the field with this lower cell gap $d$,
the fingers narrow more [Fig. 2(b)]. 

In order to decrease $B_0$, we kept this lower cell gap
and increased the injection pressure up to $p_e=22mbar$. 
$B_0$ must have been actually lowered,
since the interfaces obtained were much
noisier, and the fingers growing perpendicular to the groove even tip-split,
as reported in the previous section.
(The amount of noise necessary for a finger to tip-split is known to decrease
with decreasing dimensionless surface tension $B_0$ \cite{bensimon}). 
However, also as explained in the previous section, there was no
significant width change. The fact that a change by a factor 4.4 in 
the dimensionless surface tension $B_0$
when increasing the injection pressure from $p_e=5mbar$ up to $p_e=22mbar$
causes no visible width change implies that the mentioned change of 20-50\% in 
$B_0$ through the change in the surface tension $\sigma$ measured for other 
liquid crystals cannot cause it either.
We are therefore led to conclude that it
is the anisotropy in the surface tension and not the
dimensionless surface tension itself what accounts for the observed width
change.

Once we have understood how the introduction of an
electric field affects the width and velocity
of the fingers, 
we are in a position to explain the experimental
observations when the electric field is periodically switched on and off.
One would be tempted to understand the lateral oscillations in Figs. 1(c),
1(d) and 4(c) as standard side branches, i.e., 
due to the amplification of perturbations originating on
the tip of the fingers.
One could thus think that the periodic change in some control parameter when
switching on and off the field provided the necessary local 
perturbation on the tips to induce side branching, 
or that a large enough perturbation
due to the natural noise was further amplified through a resonance phenomenon 
with the frequency of change of this parameter 
and thus also produced visible and regular side branches, as seen
in related problems \cite{borzs99}.
In this case, the relevant
control parameter should be the anisotropy $\alpha$, since the viscosity only 
enters the time scale, and can thus not affect the shape of the pattern.
In this scenario, the fact that Fig. 1(e) shows no oscillations would be 
interpreted as the result of being too 
far from the resonance frequency, and the velocity oscillations seen in Fig. 2 
would be those sometimes associated with side branching. 

However, the velocity
turns out to decrease when the finger tip narrows, as opposite to the 
usual case. This smaller velocity of thinner tips can only be explained by
the change in the time scale due to the change in the viscosity.
Actually, we visually observe the velocity to change instantly each time
the field is switched on and off in each period, and it
roughly attains in each semiperiod when the field is on (off) the same value
than in a finger grown all the time with $E$ on (off), 
as explained in the previous section. 

The finger tips are also observed to narrow at the very moment the
field is switched on and to widen at the moment it is switched off, and the
minima (maxima) of the lateral undulations also have approximately the
same width than the finger with $E$ kept on (off), as also explained above. 

All this suggests that the lateral undulations are the 
wake left by a tip
quickly and alternately relaxing to the two different selected radii
corresponding to the two different values
(one for $E$ on and one for $E$ off) of the relevant control parameter, 
the groove anisotropy $\alpha$.

Thus, the obtained undulated fingers when periodically switching on and off
the field of Figs. 1(c) and 1(d) themselves
are the result of alternately 
relaxing between the thinner [Fig. 1(b)] and wider [Fig. 1(a)] fingers grown
all the time with or without the field, respectively. 
With this explanation, the absence of significant lateral undulations for 
too high frequencies is due to the lack of time for
the finger to relax to any of the two widths within each period, 
which should result in an intermediate width, 
as is indeed the case in Fig. 1(e). 

This mechanism seems to be decoupled from `natural' (noise-induced) side 
branching, since, when this natural side branching is already present with a
field kept off and on [Figs. 4(a) and 4(b), respectively], periodically 
switching on and off the field seems just to superimpose 
the mentioned wake of tip radius changes, but not to eliminate or 
regularize the previously present modes [Fig. 4(c)]. The fact that the two
effects be decoupled supports the idea that the width and velocity
oscillations observed when switching on and off the field 
are the result of the relaxation back and forth
between two different stationary widths and velocities, rather than that of
the amplification of perturbations coming from the tip. 

Indeed, {\em instantly} switching on and off the field with a certain period 
does not introduce any extra time scale nor control parameter in the dynamics 
{\em of each semiperiod} during which the field is either on or off. 
Each semiperiod can be understood as the relaxation with certain values of the
control parameters towards a new steady state from a given 
initial condition. Just that this initial condition turns out to be a state 
more or less close to the steady state corresponding to different values of the
control parameters. Since no extra time scale is introduced, no coupling with 
the natural noise was to be expected.

However, varying the amplitude of the field with say a sinusoidal wave instead 
of a square one would introduce as new time scale the period of the wave, so
that the dynamics would change. The injection pressure in viscous fingering 
in a channel has indeed been varied with such a sinusoidal wave. 
The interesting point is that the tip velocity follows the pressure 
modulation and the pattern obtained also displays lateral undulations of 
limited amplitude, which are symmetrical as long as no external 
element breaks this symmetry (case of two parallel grooves) \cite{rabaud}.
Also a bubble on the tip of a finger can induce the tip curvature
to oscillate periodically and give rise to symmetrical lateral undulations of
a well-defined amplitude (in the channel geometry) and periodicity. 
The amplitude is such that the outer 
envelop of the wave is a larger Saffman--Taylor finger in the channel, 
and the periodicity is
correlated to the frequency of oscillation of the tip 
\cite{couder1,couder3,rabaud}.

All these observations with pressure modulation or bubbles match our own
observations with electric field modulation, so that the mechanism of
successive relaxations between two different steady states which we propose 
might also be relevant to these other experiments. 
Our case, however, is particularly
clear thanks to the use of a square wave to modulate the electric field. 

In conclusion, two different limiting cases seem to lead to the formation of
lateral waves: (i) The amplification of small perturbations when advected
from the tip to the sides of the finger (e.g. natural, noise-induced side 
branching). (ii) Successive and alternate 
relaxations between two different finger widths, also advected from the
tip to the sides, when for some reason the tip curvature oscillates (e.g. 
periodic, instant changes in a control parameter affecting selection as in
our experiments). Of course, the lateral undulations caused by the successive 
relaxations (ii) might also be damped or 
amplified as in (i), and it can be difficult to tell whether a particular
deformation of the tip is rather a small perturbation (i) or an overall 
curvature change (ii), so that we feel that both mechanisms should be regarded 
as complementary, and experiments were a large perturbation is used to force
the dynamics might be expected to be mixed cases.

\section{Numerical results and discussion}
\label{secnumerical}

The difficulty to check the explanation of the lateral undulations
in terms of successive changes in the selected width proposed in the previous 
section lies in the fact that the finger width is not well defined.
The sides of (anisotropic) viscous fingers in the radial geometry are not 
parallel, and, most importantly, anisotropic fingers do not reach 
a steady tip radius nor velocity. There is indeed a
selection mechanism, but the first keeps growing and the latter decreasing
with time (see Ref. \cite{almgren}). Therefore, it is especially useful to 
perform numerical simulations of 
anisotropic fingers in the channel geometry to check out this scenario, 
since their sides are parallel, and, above all, their
widths and velocities do saturate and are easy to compare with one another.
Note that the experiments in the channel geometry with pressure modulation or
bubbles mentioned in the previous section \cite{rabaud} are not clear enough 
for that purpose, since the selected width keeps changing all the time as the 
effective control parameters should oscillate sinusoidally in response to a 
sinusoidal pressure or bubble forcing. In contrast, we will instantly change 
the value of the relevant control parameter (the anisotropy due to the groove
$\alpha$) in our simulations to mimic the switching on and off of the electric 
field.

On the other hand, it is well known that a thinner finger grows faster in 
dimensionless
time, although the experimental observation is just the opposite in real time.
This means 
that the change in the time scale due to the change in the viscosity 
(3.7 times larger with $E$ on)
must
be dominant over the change in dimensionless time.  
The question is whether there actually exists a range of change of the groove 
anisotropy $\alpha$ which yields
the observed narrowing of the finger 
but also respects the fact that thinner fingers grow slower in real time.

To answer this question and check the proposed explanation of the lateral 
undulations, we numerically integrate the described theoretical
model, but we run it in the channel geometry.
We use the phase-field model for
viscous fingering presented and tested in Ref. \cite{pf}. 
The only change in the model is that we now use the 
anisotropic surface tension given 
by Eq. (\ref{eq:anisotropy}). We recall the model,
\begin{equation}
\label{eq:sf}
\tilde{\epsilon} \frac{\partial\psi}{\partial t}=\nabla^2\psi+c\vec \nabla \cdot
(\theta \vec \nabla \psi)+\frac{1}{\epsilon} \frac{1}{2\sqrt 2}
\gamma(\theta )(1-\theta^2)
\end{equation}
\begin{equation}
\label{eq:pf}
\epsilon^2 \frac{\partial \theta}{\partial t}=f(\theta)+\epsilon^2\nabla^2\theta
+\epsilon^2 \kappa(\theta ) |\vec \nabla \theta |+\epsilon^2
\hat z \cdot
(\vec \nabla \psi \times \vec \nabla \theta),
\end{equation}
where $\psi$ is the stream function, $\theta$ is the phase field, 
$c\equiv (\mu -\mu_0)/(\mu +\mu_0)$ is
the viscosity contrast ($\mu$, $\mu_0$ are the viscosities of the liquid
crystal and the air, respectively)
and
$\epsilon$, $\tilde{\epsilon}$ are model parameters which must be small to
recover the sharp-interface equations of the theoretical model.
We have defined
$f(\theta )\equiv \theta (1-\theta^2)$, and
$\gamma(\theta)/2 \equiv \hat s(\theta)\cdot\{\vec\nabla
[B(\theta)\kappa(\theta)]
+\hat y\}$,
$\kappa(\theta)\equiv -\vec\nabla \cdot \hat r(\theta)$,
with $B(\theta) \equiv B[\phi=\arccos \hat y\cdot\hat r(\theta)]$,
$\hat r(\theta)\equiv \vec\nabla \theta/|\vec\nabla \theta|$
and $\hat s(\theta)\equiv \hat r(\theta) \times \hat
z$.
All quantities are dimensionless and, in particular, lengths are in units of
the channel width ($y$ is length along the channel, $x$ across it, $\hat z$
is perpendicular to the plates,
and $\phi$ is reinterpreted as the angle between $\hat y$ and the normal to
the interface).

We set $B_0=10^{-2}$, which we know to allow stable fingers for the amount of
numerical noise present even for vanishing anisotropy \cite{lc}. 
We use two different
values of the 
anisotropy, $\alpha=0.9$ and $\alpha=0.1$, to account for the cases with and
without electric field, respectively. The higher anisotropy gives the lowest
$B$ at the finger tip which we will need to resolve, and thus the value of
the interface width to use, $\epsilon=0.00625$. As for the viscosity contrast,
for numerical convenience we use $c=0.9$, 
which is known (see, e.g., Ref. \cite{lc})
to be sufficiently close to the
high viscosity contrast limit $c=1$
of the experiments. $\tilde{\epsilon}=0.4$, which suffices to resolve the
displacement of the liquid crystal by the air.
The initial condition is a cosine wave of wavelength and amplitude 1 
(the channel width) in all cases. 

Since the simulations use dimensionless variables,
the effect of the different time scale with or without the electric field does
not show up. To make it apparent, we introduce another dimensionless time
increment
\begin{equation}
\label{eq:realtime}
\Delta t' \equiv \left\{ \begin{array}{ll}
                 \Delta t & \mbox{when field is off}\\
                 a \Delta t &
                            \mbox{when field is on},
                 \end{array}
         \right.
\end{equation}
where $a=3.7$ is the measured ratio of the time scale with $E$ on and that
with $E$ off,
and we compare runs at a same time $t'$. In this way we compare runs
which would have taken the same time in the experiments, since the factor
restoring the dimensions is now the same independently of how much time was 
the field on or off during each run. Also, the phase-field equations 
[(\ref{eq:sf}) and (\ref{eq:pf})] are in the reference frame moving with the
mean interface. Since the experimental figures are in the lab frame, 
the numerical simulations (Figs. 5-7) have been
translated into the latter for comparison.

Figs. 5 and 7 are the computational, channel analogues of 
Figs. 1 and 2, obtained from experiments in the radial geometry.
In Fig. 5(a) we show a wider ($\alpha=0.1$, field off, $\lambda=0.588$)
and a thinner ($\alpha=0.9$, field on, $\lambda=0.387$) finger, 
both at $t'=7.8$,
where $\lambda$ is the finger width. 
We can see that the wider finger
does go faster in real time even for this significant change in width. 
Therefore, we conclude that a simultaneous increase in the surface tension
anisotropy 
and the viscosity, does actually explain the fact that
fingers are both narrower and slower.

In Figs. 5(b) and 5(c), interfaces are shown exactly each time the anisotropy 
was changed between the two different values in Fig. 5(a) 
(each time the field was switched on or off), which was done with a very
similar filling coefficient than in the experiments, $\xi=0.67$. 
This visually leaves no doubt of the fact that the two different widths in 
Fig. 5(a) are successively selected at the tip
to produce the pattern in Fig. 5(b).
The small mismatch between the
tails of the two front interfaces in Fig. 5(b) is presumably due to the fact 
that the viscosity contrast is not strictly 1 ($c=0.9$), so that the dynamics
in the tail region is not completely frozen.
In Fig. 5(c) the width has no time to relax to any of the two in Fig. 5(a),
and gently oscillates in the intermediate range $0.526<\lambda<0.537$.
However, the curvature seems to relax more quickly. 

%%%%%%%%%%%%%%%%%%%%%%%%%% FIG. 5 %%%%%%%%%%%%%%%%%%%%%%%%%%%%%%%%%%%%%

These relaxation processes can be seen in more
detail in Fig. 6, where we have monitored the finger width one unit length
behind the tip (which is only slightly below the asymptotic width) and the tip 
radius. The latter is the inverse of the curvature modulus of the 
zero level-set of the phase field ($1/|\kappa(\theta=0)|$), and therefore
of the interface. To avoid spurious lattice oscillations, such a radius is
plotted only when the finger tip hits near a grid point 
($|\theta|<10^{-3}$ at the tip of the finger). The solid and dashed lines
correspond to the runs in Fig. 5 (b) and 5 (c) respectively. In the case
of the lower frequency (solid lines), we can see that the tip radius 
relaxes always first to its asymptotic value, and is then followed by the 
finger width as the information of the tip is left behind.
Thus, for the higher frequency (dashed lines), the
finger has not enough time to relax to its asymptotic widths, but the curvature
almost attains its asymptotic values.
On the other hand, for the lower frequency it is possible to
observe that the tip widens
much faster than it narrows, as can be seen both in the tip width and its
radius, but especially in the first. This behavior
may be expected in connection with the existence of a set of (unstable)
solutions with larger width than the selected one, whose proximity may
effectively slow down the relaxation dynamics.

%%%%%%%%%%%%%%%%%%%%%%%%%%% FIG. 6 %%%%%%%%%%%%%%%%%%%%%%%%%%%%%%%%%

Finally, in Fig. 7 we show the evolution of the tip position for the runs in
Fig. 5. The steeper (less steep) straight, long-dashed line corresponds to 
the wider (thinner) finger in Fig. 5 (a), i.e. to the case with the field 
off (on). 
The initial relaxation to the stationary 
velocity is so fast that it is almost invisible at this scale.
The runs in Figs. 5 (b) and (c) correspond to the solid and dotted lines in 
between, respectively. We can see that, for the lower frequency (solid line), 
the velocity successively relaxes to the values with or without field of the
straight, long-dashed lines. Initially, however, it attains a value slightly 
below (above) the steady velocity when it relaxes to a lower (higher) velocity.
This effect is more apparent for the relaxation to a lower velocity. 
In contrast, for the higher frequency (dotted line), we are left with these
slightly too low or high initial values of the velocity, 
since the field is switched on or off again just when the 
velocity was about to achieve its asymptotic value. This is quite similar to
what happened to the curvature for the higher frequency (lower dashed line in
Fig. 6), in contrast with the failure of the finger width to relax (upper
dashed line in Fig. 6). So the finger velocity seems to be more correlated
to the tip curvature than to the finger width.

%%%%%%%%%%%%%%%%%%%%%%%%%%%%%% FIG. 7 %%%%%%%%%%%%%%%%%%%%%%%%%%%%%%%%%%%%

In order to compare with the experiments of 
Rabaud {\it et al.} with two opposite grooves in the channel geometry 
in which they modulated the injection pressure \cite{rabaud},
we have repeated our simulations changing the
dimensionless surface tension from $B_0=10^{-2}$
to $B_0=6.5\times 10^{-4}$ (with $\epsilon=0.005$) 
and keeping its anisotropy to $\alpha=1$. Note that, indeed, an instant change 
in pressure is equivalent to a change in the time scale and the dimensionless
surface tension. In the experiments of Rabaud {\it et al.} 
the modulation was sinusoidal,
which also introduces an extra time scale, but they nevertheless obtained
symmetrical lateral waves of limited amplitude as ours \cite{rabaud}. 
In our simulations, 
we use instant changes in the dimensionless surface tension, and we obtain
qualitatively the same results than in Figs. 5 and 7. The instant changes
make the saturation of the finger width possible, and the fact that the finger
width does saturate to the values with a constant $B_0$ suggests that the
basic mechanism for the lateral waves observed by Rabaud {\it et al.}
when modulating the pressure
might also be the relaxation towards two different steady widths. 

\section{Conclusions}
\label{secconclusions}

We have performed viscous fingering experiments in a radial Hele--Shaw cell,
where the more viscous
fluid was a liquid crystal mixture in its nematic phase.
After ruling a single groove across the center of the cell, we achieved stable 
finger tips in the direction of the groove (otherwise unstable).
By applying an electric field perpendicular to the cell, we oriented the
nematic director in this direction, 
which resulted in thinner and slower fingers.
We then periodically switched on and off the field to find oscillations in the
finger width and velocity, with an amplitude which decreased
as the switching frequency was increased. 

We explain how fingers are slower when the field is
on because the viscosity of the liquid crystal is higher with the director
perpendicular to the cell, and that the reason why
they are thinner may be attributed to an increase in the anisotropy due to the
groove when the field is on. Also the surface tension is reduced when the 
field is switched on, but it cannot affect so strongly the
finger width, since no significant width change was observed by increasing 
the excess pressure, and both a decrease in the surface tension and
an increase in the excess pressure would lower the
dimensionless surface tension. 
The proposed scenario reproduces the experimental observations, as
shown by numerical integration in the
channel geometry of a simplified theoretical model.
We also explain the finger width and velocity
oscillations as the result of the relaxation back and forth between the
selected tip radii and
velocities with the field on and off, as suggested by the experiments and
clearly seen in the numerical integration of the theoretical model.

We discuss how this latter result might be relevant to experiments 
with a bubble on the tip of a finger and especially when modulating the
injection pressure in a channel with two parallel grooves 
\cite{couder1,couder3,rabaud}. 
We reproduce the qualitative observation that the lateral
waves are symmetric and of limited amplitude for such a pressure modulation 
by simulations instantly changing the dimensionless surface tension.
We point out that the amplification of small tip perturbations describing
natural, noise-induced side branching, and the successive relaxations between
to steady widths describing the formation of lateral undulations when
periodically changing a control parameter seem to be two complementary 
mechanisms for lateral wave formation, and that experiments forcing the 
dynamics with large perturbations might be understood as mixed cases.

\acknowledgements{
We acknowledge financial support from the Direcci\'on
General de Ense\~{n}anza Superior (Spain), Projects No. BXX2000-0638-C02-02 and
BFM2000-0628-C03-01, from the National Scientific Research  
Foundation (Hungary), Grants Nos. OTKA F022771 and OTKA T031808, 
and the European Commission, Project Nos. ERB FMRX-CT96-0085.
Simulations have been carried out using the resources at CESCA and
CEPBA, coordinated by $\rm C^4$.
R.F. also acknowledges a grant from the 
Comissionat per a Universitats i Recerca (Generalitat de Catalunya).

\newpage
\begin{figure}
\centerline{\psfig{file=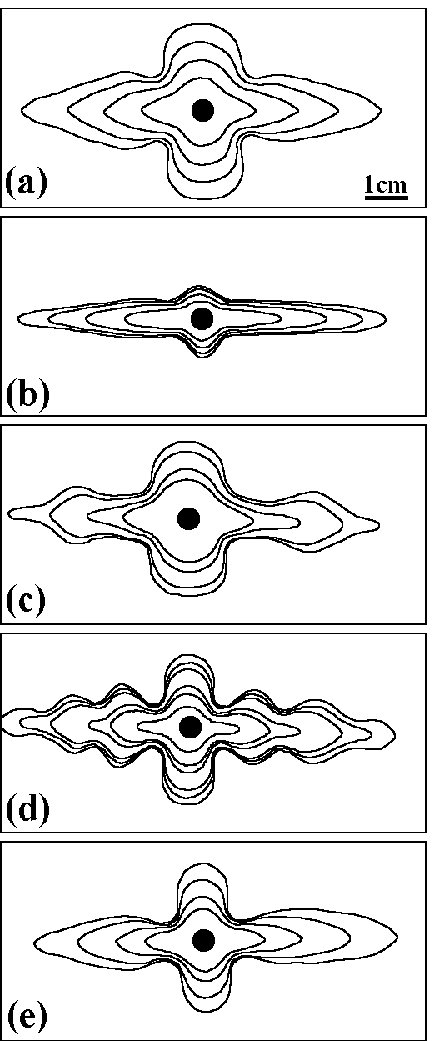,width=.47\textwidth}}
\vskip 1cm
\caption{
Air -- nematic interfaces at subsequent times. $d=320\mu m$, $p_e = 5mbar$. 
(a) $E$ off, $t=0.32s$, $0.64s$, $0.96s$, $1.32s$; 
(b) $E=0.32V/\mu m$ on, $t=1.08s$, $2.20s$, $3.32s$, $4.44s$; 
(c)-(e) modulated $E$, $\xi = 0.68$:
(c) 
$\nu=0.667Hz$, $t=0.4s$, $1.44s$, $1.84s$, $2.64s$; 
(d) $\nu=1.01Hz$, $t=0.72s$, $1.04s$, $1.68s$, $2.04s$, $2.68s$, $2.88s$; 
(e) $\nu=4.55Hz$, $t=0.52s$, $1.04s$, $1.56s$, $2.08s$.  
Subfigures (c) and (d) show the interfaces each time $E$ was 
switched on/off.}
\label{fig1}
\end{figure}

\newpage
\begin{figure}
\centerline{\psfig{file=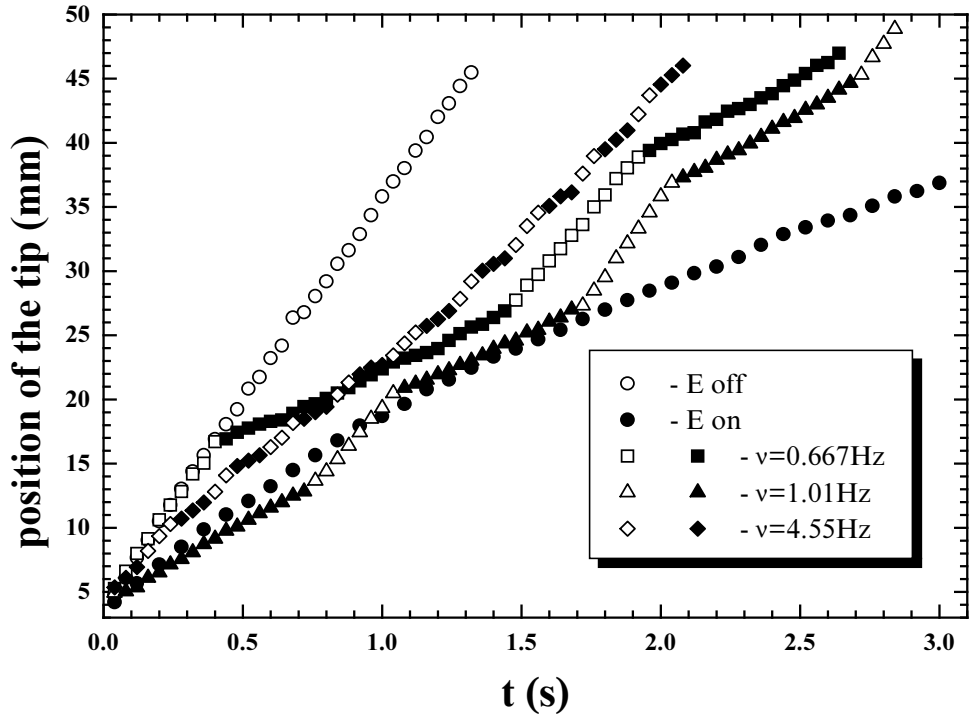,width=\textwidth}}
\caption{
Position of the tip of the main fingers in Fig. 1 vs. time.
Filled (empty) symbols denote $E$ on (off). Circles correspond to the
experiments with field kept
off (empty) of Fig. 1(a) or on (filled) of Fig. 1(b), whereas the other
symbols stand for the different frequencies with which the field was switched
on/off:
squares, triangles, and diamonds for Figs. 1(c), 1(d), and 1(e), respectively.
}
\label{fig2}
\end{figure}

\newpage
\begin{figure}
\centerline{\psfig{file=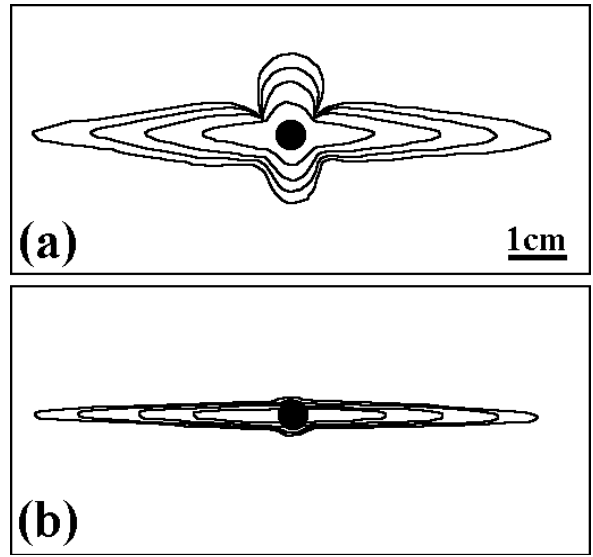,width=.47\textwidth}}
\vskip 1cm
\caption{
Same than Fig. 1, but $d=190\mu m$. 
(a) $E$ off, $t=0.24s$, $0.60s$, $1.00s$, $1.40s$; 
(b) $E= 0.55V/\mu m$ on, $t=0.60s$, $1.12s$, $1.72s$, $2.24s$.
}
\label{fig3}
\end{figure}

\newpage
\begin{figure}
\centerline{\psfig{file=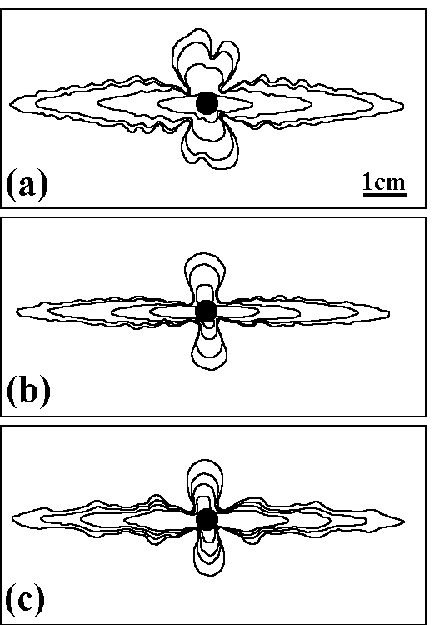,width=.47\textwidth}}
\vskip 1cm
\caption{
Same than Fig. 3, but $p_e = 22mbar$. 
(a) $E$ off, $t=0.04s$, $0.12s$, $0.20s$, $0.24s$; 
(b) $E= 0.55V/\mu m$ on, $t=0.08s$, $0.16s$, $0.28s$, $0.36s$; 
(c) modulated $E= 0.58V/\mu m$, $\nu=8.42Hz$, $\xi = 0.68$, 
$t=0.12s$, $0.20s$, $0.28s$, $0.36s$.
}
\label{fig4}
\end{figure}

\newpage
\begin{figure}
\centerline{\psfig{file=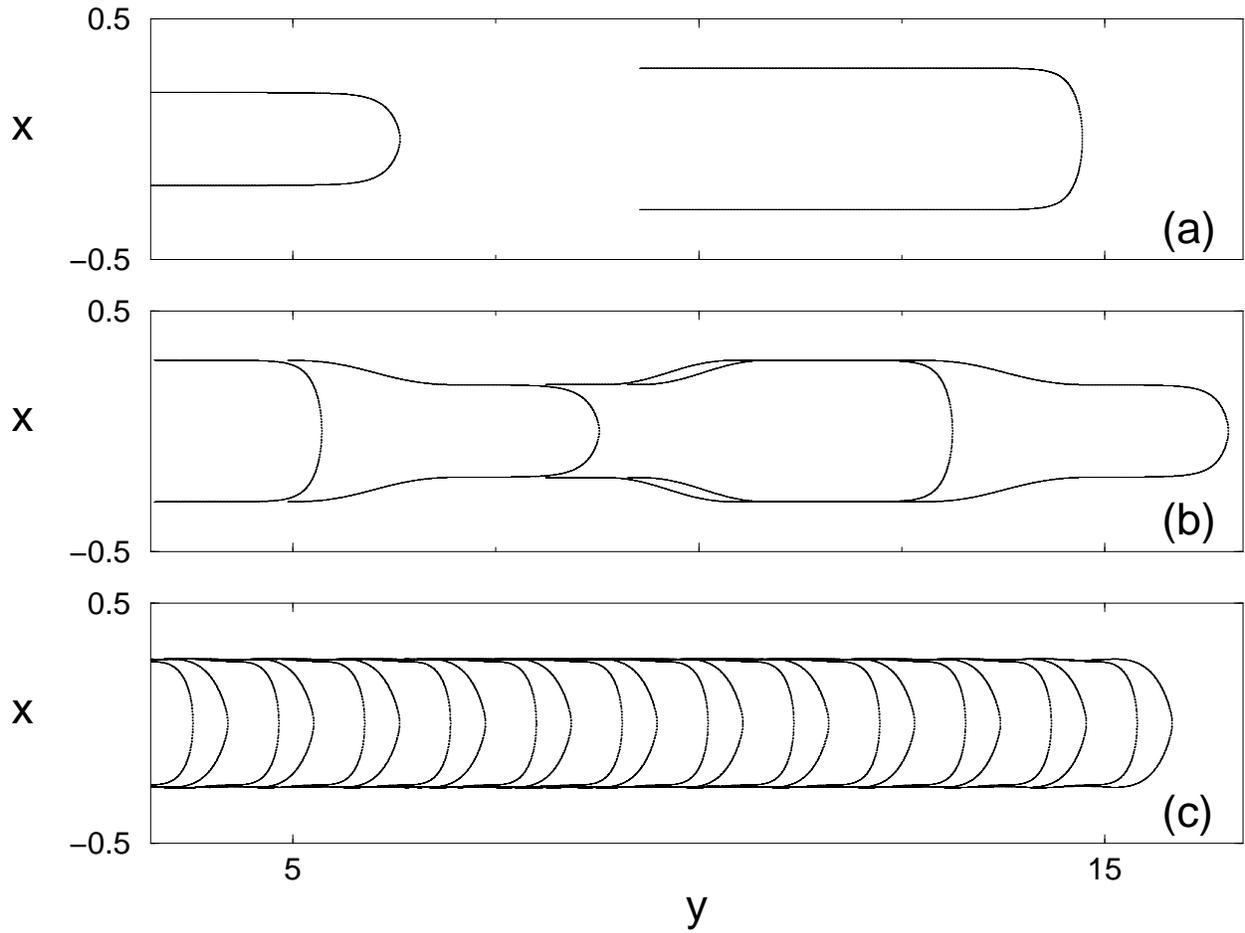,width=\textwidth}}
\vskip 1cm
\caption{
Interfaces in the channel geometry, simulated in the reference frame
moving with the mean interface and then translated into the lab frame. 
$B_0=10^{-2}$, $\epsilon=0.00625$, $c=0.9$, $\tilde{\epsilon}=0.4$.
(a) Change of width in the stationary pattern at $t'=7.8$ when changing
from $\alpha=0.1$ (wider finger) to $\alpha=0.9$ (thinner finger).
(b),(c) Periodic, instantaneous switch of $\alpha$ between the two
values in (a), with a lower (b) and a higher (c) frequency.
Interfaces are shown each time the field 
is switched on or off with $\xi=0.67$, until $t'=15.3$.
}
\label{fig5}
\end{figure}

\newpage
\begin{figure}
\centerline{\psfig{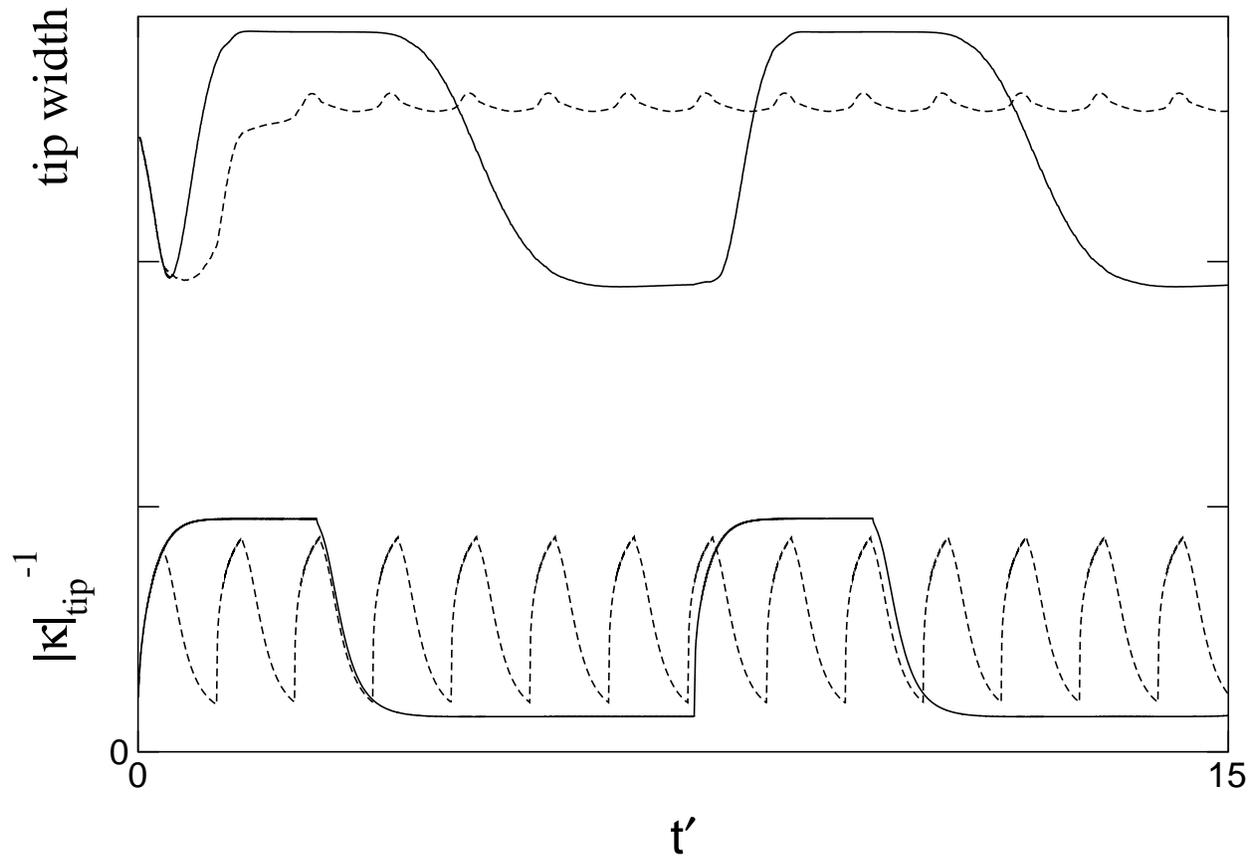}}
\vskip 1cm
\caption{
Finger width at one unit length behind the tip and tip radius
vs. rescaled time ($t'$) 
for the runs in Figs. 5(b) (solid lines) and 5(c) (dashed lines). Recall that
the unit length is the channel width.
}
\label{fig6}
\end{figure}

\newpage
\begin{figure}
\centerline{\psfig{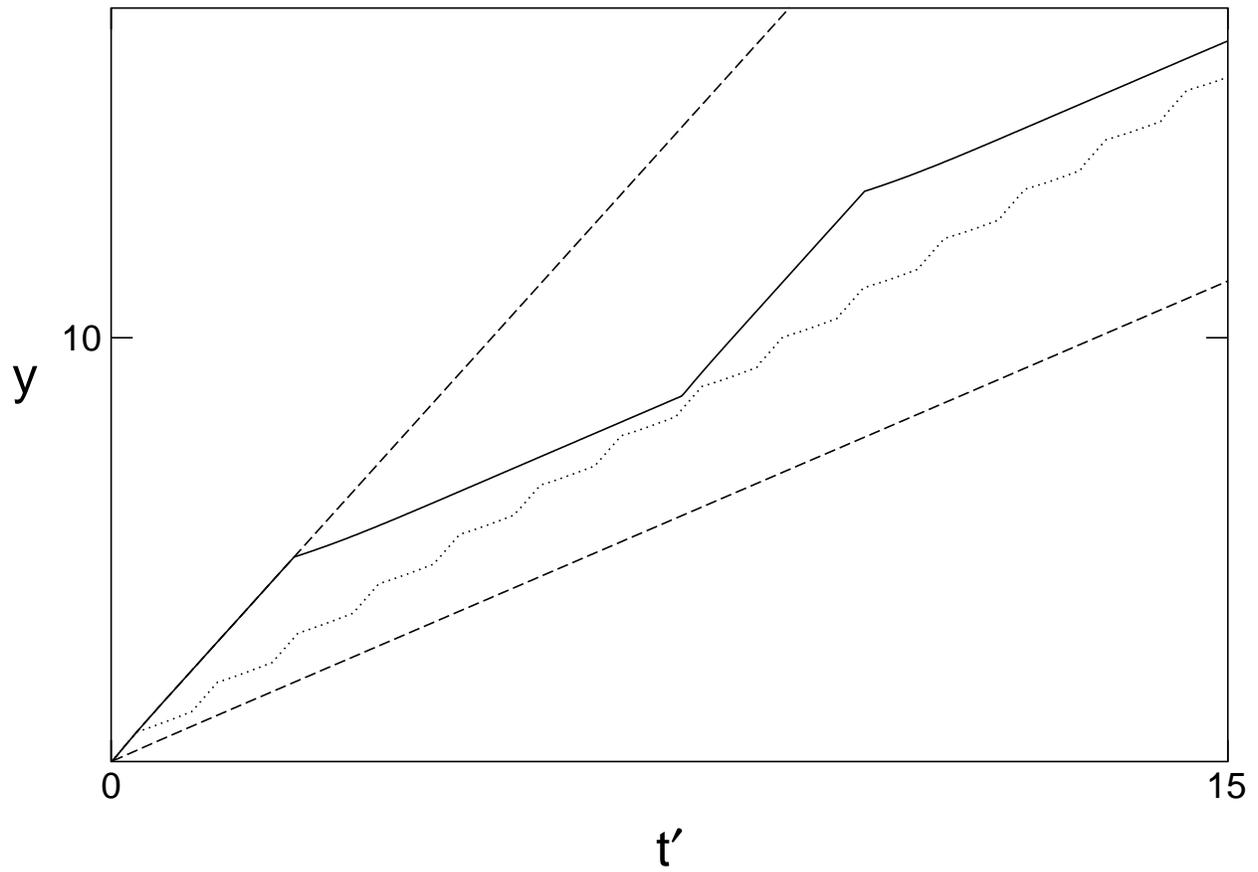}}
\vskip 1cm
\caption{
Tip position ($y$) vs. rescaled time ($t'$) for the runs in Fig. 5.
The steeper (less steep) straight, long-dashed line corresponds to the
wider (thinner) finger in Fig. 5(a). The solid and dotted lines in between
correspond to the runs in Figs. 5(b) and 5(c) respectively.
}
\label{fig7}
\end{figure}

%\end{multicols}


\begin{references}

\bibitem{pelce} E. Ben-Jacob and H. Levine, Adv. Phys. {\bf 49}, 395 (2000);
J. P. Gollub and J. S. Langer, Rev. Mod. Phys. {\bf 71}, S396 (1999);
{\it Solids far from Equilibrium}, edited by C. Godr\`eche 
(Cambridge University Press, Cambridge, 1992);
D. A. Kessler, J. Koplik, and H. Levine, Adv. Phys. {\bf 35}, 255 (1988);
P. Pelc\'e, {\it Dynamics of Curved Fronts} (Academic, New York, 1988).

\bibitem{mccloud1} K. McCloud and J. Maher, Phys. Rep. 
{\bf 260}, 139 (1995).

\bibitem{pater1} L. Paterson, J. Fluid Mech.
{\bf 113}, 513 (1981).

\bibitem{couder1} Y. Couder, O. Cardoso, D. Dupuy, P. Tavernier and 
W. Thom, Europhys. Lett. 
{\bf 2}, 437 (1986). 

\bibitem{couder3} Y. Couder, N. Gerard and M. Rabaud, Phys. Rev. A 
{\bf 34}, 5175 (1986).  

\bibitem{benj9} E. Ben-Jacob, R. Godbey, N. Goldenfeld, J. Koplik, H. Levine, 
T. M\"uller and L. Sander, Phys. Rev. Lett.
{\bf 55}, 1315 (1985). 

\bibitem{benj4} E. Ben-Jacob and P. Garik, Nature (London)
{\bf 343}, 523 (1990). 

\bibitem{chen} J. Chen and D. Wilkinson, Phys. Rev. Lett. 
{\bf 55}, 1892 (1985); J. Chen, Experiments in Fluids 
{\bf 5}, 363 (1987). 

\bibitem{horvath1} V. Horv\'ath, T. Vicsek and J. Kert\'esz, 
Phys. Rev. A {\bf 35}, 2353 (1987). 

\bibitem{matsu1} M. Matsushita and H. Yamada, J. Cryst. Growth 
{\bf 99}, 161 (1990). 

\bibitem{buka96} see e.g. \'A. Buka, in {\it Pattern Formation in Liquid 
Crystals}, edited by \'A. Buka and L. Kramer 
(Springer, New York, 1996) p. 291

\bibitem{lc}
R. Folch, J. Casademunt and A. Hern\'andez-Machado, Phys. Rev. E 
{\bf 61}, 6632 (2000).

\bibitem{rabaud} M. Rabaud, Y. Couder and N. Gerard, Phys.Rev. A 
{\bf 37}, 935 (1988).

\bibitem{pf}
R. Folch, J. Casademunt, A. Hern\'andez-Machado and L. Ram\'{\i}rez-Piscina,
Phys. Rev. E {\bf 60}, 1724 (1999);
{\it ibid.} {\bf 60}, 1734 (1999).

\bibitem{borzs99} T. B\"orzs\"onyi, T. T\'oth-Katona, \'A. Buka and 
L. Gr\'an\'asy, Phys.Rev.Lett. {\bf 83}, 2853 (1999). 

\bibitem{pleiner1} H. Pleiner and H. Brand, in 
{\it Pattern Formation in Liquid Crystals}, edited by \'A. Buka and 
L. Kramer (Springer, New York, 1996) p.15

\bibitem{dubois1} E. Dubois-Violette and P. Manneville, in 
{\it Pattern Formation in Liquid Crystals}, edited by \'A. Buka and 
L. Kramer (Springer, New York, 1996) p.91

\bibitem{toth00} T. T\'oth-Katona and \'A. Buka (unpublished).

\bibitem{benjacob}
E. Ben-Jacob, P. Garik, T. Mueller and D. Grier, Phys. Rev. A {\bf 38}, 1370
(1988);
E. Ben-Jacob and P. Garik, Physica D {\bf 38}, 16 (1989).

\bibitem{tsvet1}
V.A. Tsvetkov, O.V. Tsvetkov and V.A. Balandin, Mol. Cryst. Liq. Cryst. Sci. 
Technol., Sect. A {\bf 329}, 305 (1999). 

\bibitem{bensimon}
D. Bensimon, Phys. Rev. A {\bf 33}, 1302 (1986).

\bibitem{almgren}
R. Almgren, W.-S. Dai and V. Hakim,
Phys. Rev. Lett. {\bf 71}, 3461 (1993)

\end{references}
\end{document}